\documentclass[conference]{IEEEtran}
\IEEEoverridecommandlockouts
\usepackage{cite}
\usepackage{amsmath,amssymb,amsfonts}
\usepackage{booktabs, makecell}
\usepackage{algorithmic}
\usepackage{graphicx}
\usepackage{textcomp}
\usepackage[nolist]{acronym}
\usepackage{breqn}
\usepackage[T1]{fontenc}
\usepackage{url}
\usepackage{xcolor}
\usepackage{silence}
\usepackage[textsize=tiny,colorinlistoftodos]{todonotes}
\makeatletter
\define@key{todonotes}{bh}[]{
	\setkeys{todonotes}{author=\textbf{Bin}, color=yellow!30}}%
\define@key{todonotes}{ak}[]{
	\setkeys{todonotes}{author=\textbf{Anthony}, color=blue!30}}%
\makeatother
\usepackage[normalem]{ulem}
\newcommand{\revise}[2]{{\color{red}\sout{#1}}{\color{blue}#2}}

\newif\ifreviewmode
\reviewmodetrue 
\reviewmodefalse

\ifreviewmode
\else
  \renewcommand{\todo}[1]{} 
  \renewcommand{\revise}[2]{#2} 
\fi

\def\BibTeX{{\rm B\kern-.05em{\sc i\kern-.025em b}\kern-.08em
    T\kern-.1667em\lower.7ex\hbox{E}\kern-.125emX}}
\begin{document}
\title{Resource Allocation in Mobile Networks:\\A Decision Model Of Jockeying in Queues
}

\author{
	\IEEEauthorblockN{Anthony~Kiggundu\IEEEauthorrefmark{1},~Bin~Han\IEEEauthorrefmark{2},~Dennis~Krummacker\IEEEauthorrefmark{1}~and~Hans~D.~Schotten\IEEEauthorrefmark{1}\IEEEauthorrefmark{2}}
	\IEEEauthorblockA{
		\IEEEauthorrefmark{1}German Research Center for Artificial Intelligence (DFKI), Germany\\
		\IEEEauthorrefmark{2}University of Kaiserslautern (RPTU), Germany\\
	}
}

\maketitle

\begin{abstract}
\revise{Application specific network slices offered by different vendors is one of those approaches proposed to bridge the gap between the demand and supply of resources in next generation communication networks.}{Use-case-specific network slicing in decentralized multi-tenancy cloud environments is a promising approach to bridge the gap between the demand and supply of resources in next-generation communication networks.\todo{C1}}
Our findings associate different slice \revise{configurations}{profiles} to queues in a multi-\revise{vendor}{server} setting\revise{}{,} such that \revise{consumers}{tenants\todo{C2}} continuously assess their preferences and make rational decisions \revise{that guide the optimal usage of the resource pools}{to minimize the queuing delay\todo{C3}}.

\revise{In deviation}{Deviated\todo{C4}} from classical approaches that \revise{model for jockeying using statistical techniques}{statistically model the jockeying phenomena in queuing systems}, our work \revise{is pioneer in the class of behavioral models that seek to characterize for the impatient customer's preferences}{pioneers to setup a behavioral model of jockeying impatient tenants}. \revise{This is so as to break away from the centralized control in multi-queue systems and shifting towards a more decentralized decision making mechanism such that, the decision to jockey is premised on measures about the continuous sensitivity to variations in queue parameters.}{This will serve as a basis for decentralized management of multi-queue systems, where the decision to jockey is individually made by each tenant upon its up-to-date assessment of expected waiting time.\todo{C5}}

Additionally, \revise{our open-source developmental tooling provides a test-bed for unraveling underlying parametric dependencies in similar Monte Carlo simulations. One such dependency is that between the expected waiting time and the number of times tasks are moved from one queue to another. To validate the novelty of our behavioral Monte Carlo experiment under these perturbations, we show how the dependencies can analytically be characterized for.}{we carry out numerical simulations to empirically unravel the parametric dependencies of the tenants' jockeying behavior.\todo{C6}}  
\end{abstract}

\section{Introduction}
Recent trends in resource allocation in communication systems suggest a shift in paradigm towards partitioning schemes of the shared physical network along application specific virtual networks that scale to obligatory \ac{QoS} or \ac{QoE} requirements \cite{lorawan,aloqaily2018congestion}. It is proposed that the decoupling and abstraction of functionality in lower layers as championed by the \ac{O-RAN} community with the front-haul or mid-haul (\ac{RAN}) elements (like the \ac{RU}, \ac{CU} and \ac{DU}) \cite{oran_challenges} operational as \ac{SDN} on cloud platforms will constitute different network slice configurations \cite{han2022impatient}. That, this will increase the degrees of freedom with regards to consumer preference \cite{liu2015fair, hyytia2013optimizing} and address existing multi-vendor interoperability challenges \cite{Jiang_2021}.

Efforts towards standardization of interfaces for interactivity by the \ac{3GPP} Consortium also propose the exposure of network core function metrics to accelerate the intelligence requisite for self-organising behavior. Metrics about sessions (\ac{SMF}), network slices (\ac{NSSF}), billing, subscription etc, motivate further studies that evaluate the applications of queuing theory \cite{7819347,zakaria.et.all} to routing traffic in communication networks than before \cite{razavilar2000traffic}.

We subscribe to the notion that these metrics will drive the rationale of the impatient customer waiting in a queue to renege (tasks are pushed but withdrawn before being serviced), balk (tasks are not pushed to the queue whatsoever) \cite{han2019utility, han2020multiservice} or jockey (tasks can be moved from one queue to another) regardless of the discipline (\ac{LCFS}, \ac{SIRO} etc). In the context of resource sharing, specific to \ac{5G} or \ac{6G} communication systems, the benefits jockeying will supposedly bring with regard to optimal server utilization and general system performance \cite{sns, xin2019performance} is still an open discussion given the complexity that will be introduced in these networks. Here, we present a decision model of the impatient customer's preferences as an empirical study, focused on measuring the sensitivity patterns between different queue descriptors and the frequency with which tasks are freely moved from one similar vendor slice offering to another.

The contributions of this documentation are succinctly two-fold:
\begin{itemize}    
    \item Most studies that model for the impatient tenant's behavior in multi-server systems adopt stochastic methods where jockeying is centrally controlled and premised on a preset threshold. This threshold is derived from the difference between the lengths of any two given queues. It can however be argued that, basing the jockeying phenomenon on such a threshold for task offloading use cases in \ac{MEC} systems is not plausible given the aleatory manifestations arising from the inter-dependency and latency requirements pegged to these systems. We propose a decentralized decision making approach for multi-server setups as a behavioural model such that the rationale for the preference of a specific service line is premised on the tenant's continuous evaluation of the task's expected waiting time. 
    \item \revise{}{The complexity introduced by the inter-dependencies between various descriptors bulges the state space definition making the statistical modeling intractable. To develop a model around these complexities then requires Monte Carlo simulators and } unlike existing libraries like SimPy \cite{simpy}, Simmer \cite{simmer} that have been previously deployed in both reneging and balking studies \cite{TININI2020102030, 9865799}, none exists for the experimental purposes of jockeying behavior. We open source our developmental tooling paving way for unraveling hidden phenomena in behavioral modeling of queuing systems. In essence, descriptors with their respective degrees of significance can be identified first so as to understand how variations in measures of different system parameters affect the impatient customer. One such effect is frequency of jobs switching resource pools and our numerical findings characterize for this frequency to validate the results from our Monte Carlo simulation. 
\end{itemize} 
In the methodology section we describe our computational setup and highlight some assumptions we adopted to realize a sufficient level of abstraction. A summary of results from multiple iterations is then presented, from which an expression for the fitted model is evaluated. We sum up this documentation with a brief discussion about our findings and areas worth inquest into the impatient consumer's behavior in next generation mobile communication systems.

\section{Methodology}
We prototype a setup with two infinite buffers $Q_{i}$ and $Q_{j}$ (\ac{M/M/C} (C=2)), where each new entry had the exposure to basic buffer size information such that admissions occurring at constant arrival rate $\lambda$ were driven by the "join the shorter buffer" strategy. Tasks in the queue were processed at heterogeneous exponentially distributed service rates $\mu_{i}$ and $\mu_{j}$ ($ i,j=\{1,2\}$ ) for each simulation run; That is, the service rates at each run were computed using \eqref{eqn:srv_rates} while restricting their summation ($\lambda = \mu_{i} + \mu_{j}$) to keep the dynamics in the system minimal. 
\begin{equation}
\mu_{i} = \frac{\lambda + \delta \lambda} {2}, \quad
 \mu_{j} = \frac{\lambda - \delta \lambda} {2}
 \label{eqn:srv_rates}
\end{equation}
\textit{where $\delta\lambda$ was a random parameter to guide the magnitude of $|\mu_{1} - \mu_{2}|$}.
\\
Our setup was composed with admissions that followed a Poisson distribution to join a service line, technically simulated as a Python priority queue with multiprocessing.  Tasks were processed following a \ac{FCFS} scheme at exponentially distributed service rates to yield service times vectors $T_{Q_{i}}=[ t_{i}, t_{i+1} \ldots t_{z}]$ and $T_{Q_{j}} = [ t_{j}, t_{j+1} \ldots t_{z}]$ at each iteration. 

\subsection{Waiting time as determinant metric for jockeying decisions} 
The rationale to jockey was premised on the evaluations of the arithmetic mean of the expected time a task would take until service completion, having landed at a given pose in the lane. Specific to our implementation, if at the time the switching decision was made, $\beta$ new arrivals reached the preferred buffer before the jockeying job, then the expected waiting time of the jockeying job landing in the preferred queue (for example $Q_{i}$) at a known pose $\tau = k +\beta + 1$ was computed using \eqref{eqn:eqn_tw}, otherwise evaluated from Little's Law.
\begin{equation}
 T_{w,\tau} (Q_{i})=
   \begin{cases}        
        \sum_{i=1}^{m}\frac{t_{i,\tau}}{m} & \text{if } \tau \in \mathbb{K}\\
        \frac{L_{Q_{i}}}{\lambda} & \text{if } \tau \notin \mathbb{K}
    \end{cases} 
    \label{eqn:eqn_tw}
\end{equation}
\textit{where m was the count over all $t_{i}$ entries at pose $\tau$,  $\mathbb{K}$ denoted a set of positions,  $L_{Q_{i}}$ was the number of tasks in $Q_{i}$ } 
\\ 

\begin{figure}[ht]
  \centering
  \includegraphics[width=\linewidth]{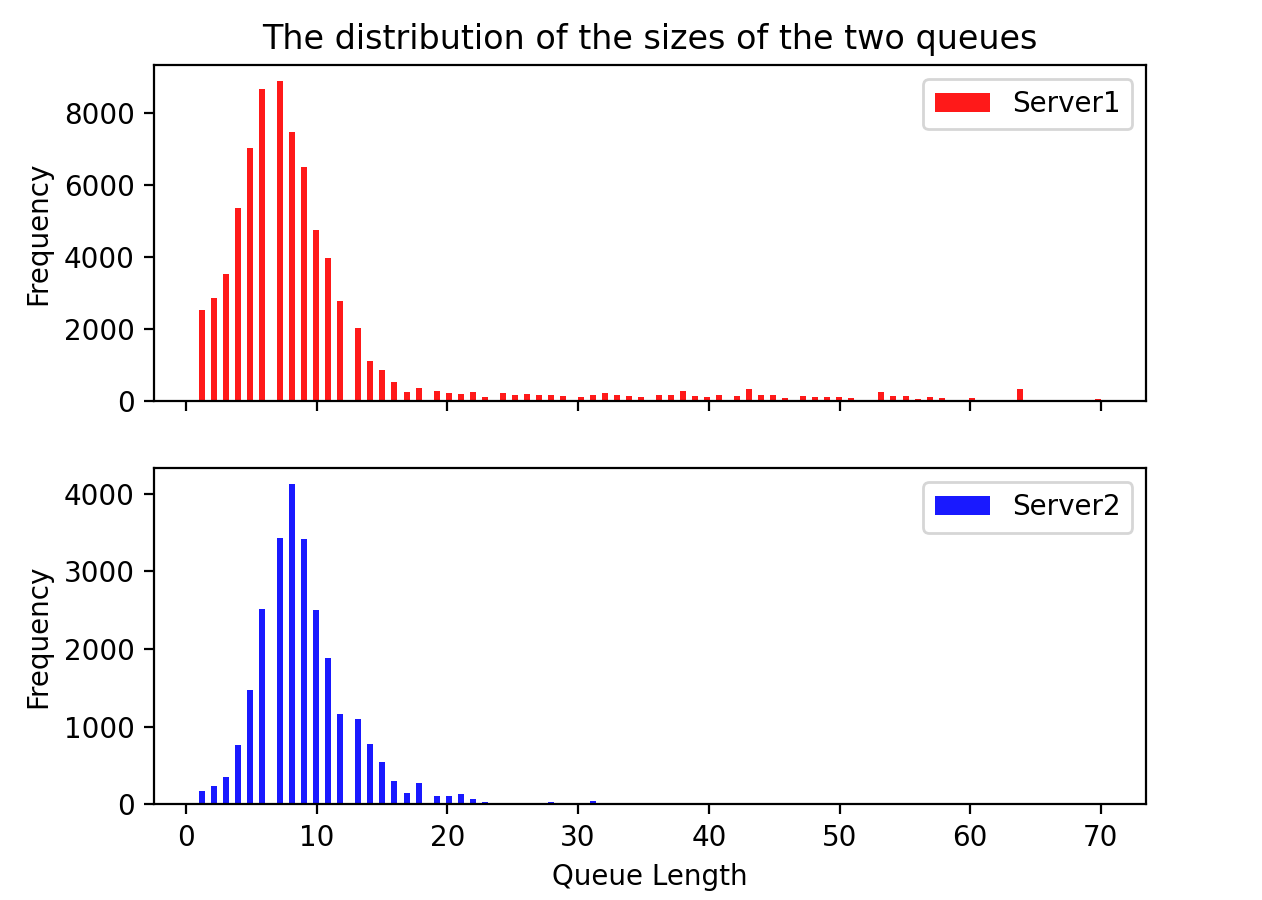}
  \caption{\footnotesize The distribution of the sizes of the buffers suggested a skewed orientation whose generalizations tended to Gaussian.}
  \label{fig:gaussian}
\end{figure}

During the entire simulation run, the rational customer reviewed the decision to jockey given a departure(s) in either queues and the probable number of new arrivals at a particular time interval $t+1$. We assumed the arrival and the departure processes followed a Poisson distribution \cite{burke}, which was theoretically foundational to the characterization for the probability that a new job was routed to a specific queue. This probability of a new job being routed to a particular queue over another then evolved from the "join the shorter queue" admission scheme. Borrowing concepts from probability theory, we determined the shorter queue by instantiating $X$ and $Y$ to denote independent random variables sampled from the distribution of the lengths of $Q_{i}$ and $Q_{j}$ respectively. 
Figure \ref{fig:gaussian} was representative of these distributions after multiple Monte Carlo iterations and it was evident that the distributions were inherent of Gaussian characteristics. 
\revise{\noindent}{}%
Given the lengths of $Q_{i}, Q_{j}$ as $L_{Q_{i}}, L_{Q_{j}}$ respectively, it could therefore be stated that $0<\beta<\lambda$ new tasks at interval $t+1$ were routed to say queue $Q_{i}$ with probability $P(X=Q_{i} < Y=Q_{j})$, conditioned on the size $\tau$ of $Q_{i}$ ($\tau_{i} = L_{Q_{i}}+\beta + 1$) at $t$ not exceeding the size of $ Q_{j}$ ($\tau_{j} = L_{Q_{j}}+\beta + 1$) and this probability we denoted as $P^{\tau}_{Q_{i,j}}(t+1)$. 

Our decision model therefore adopted Bayes' Theorem in the formulation \eqref{eqn:bayes} for the shorter queue job admission probabilities. Letting the density functions of $X$ and $Y$ be $f_{X}(x)$, $f_{Y}(y)$ respectively, then:
\begin{equation}
    P^{\tau}_{Q_{i,j}}(t+1) = P(X < Y|Y>\tau_{i}) = \frac{\int_{1}^{\tau_{i}} \int_{0}^{y} f_{X,Y}\mathrm{d}x\mathrm{d}y} {\int_{1}^{\tau_{i}}f_{Y}(y)\mathrm{d}y}
    \label{eqn:bayes}
\end{equation} 
\textit{where the definitions for $f_{Y}(y)$ or $f_{X}(x)$ evolved from \eqref{eqn:gaussian} given their inherent Gaussian properties.}
\begin{equation}
  \begin{split}
    f_X(x) = \frac{1}{\sigma\sqrt{2 \pi}} e^{-\frac{{(x-x^{\prime})}^2}{2\sigma^{2}}}, \hspace{20pt} \textrm{for all } x \in \mathbb{R}. \\  
    f_Y(y) = \frac{1}{\sigma\sqrt{2 \pi}} e^{-\frac{{(y-y^{\prime})}^2}{2\sigma^{2}}}, \hspace{20pt} \textrm{for all } y \in \mathbb{R}.
  \end{split}
    \label{eqn:gaussian}
\end{equation}

such that:
\begin{equation}
    P^{\tau}_{Q_{i,j}}(t+1) =  \frac{ \frac{1}{2\sigma\sqrt{2 \pi}} \int_{y}^{\tau_{i}} \int_{x}^{y} e^{-\bigg[\big(\frac{{x-x^{\prime}}}{{2\sigma}}\big)^{2}+ \big(\frac{{y-y^{\prime}}}{{2\sigma}}\big)^{2}\bigg]} dx dy } {\int_{y=1}^{\tau_{i}} \frac{1}{\sigma\sqrt{2 \pi}} e^{-\big(\frac{{y-y^{\prime}}}{2\sigma}\big)^{2}}dy}    
  \label{eqn:finalbayes}
\end{equation}  \textit{where $\sigma$  denoted the standard deviation, $x^{\prime}$ and $y^{\prime}$ were the mean sizes of either queues. }

\revise{\noindent}{}From \eqref{eqn:finalbayes}, the portion $\beta \leq \lambda$ of incoming tasks therefore that preferred the shorter buffer was arithmetically computed \eqref{eqn:measure_joins} as a product of the number of new jobs ($\lambda$) joining the queue at $t+1$ and the probability that all new jobs joined a particular buffer.
\begin{equation}
  \beta = \lambda * P^{\tau}_{Q_{i,j}}(t+1)  
  \label{eqn:measure_joins}
\end{equation} 

\revise{\noindent}{}The decision to move tasks to an alternative queue however, was composed based on the cases defined by \eqref{eqn:arrival_choices}; where at any iteration step, a move was influenced by the probability of $\beta \leq \lambda$ new arrivals at $t+1$ being pushed to a given buffer at the instantaneous $N \geq 1$ departure(s) from the queue or both events happening simultaneously.
\begin{equation}    
    F^{\overrightarrow{ij}}_{T_{w}| \tau}(t_{w}|\tau) =
    \begin{cases}       
       T_{w} \quad | \quad P^{\beta}_{Q_{i,j}}(t+1) \quad  & \text{if } \beta,N \geq 1\\
       T_{w} \quad | \quad P(N\geq 1) \qquad & \text{if } \beta = 0  \\
       T_{w} \qquad \qquad \qquad & \text{if } \beta, n = 0 
    \end{cases}
    \label{eqn:arrival_choices}
\end{equation}
\textit{where $T_{w} = \mathbb{E} \bigl[T_{w,k}(Q_{i})\bigr] > \mathbb{E} \bigl[T_{w,\tau}(Q_{j})\bigr]$, $\tau_{i} = L_{Q_{j}}+\beta+1$} 

\revise{\noindent}{}Since it was known that at $t+1$, the arrivals $\lambda$ obeyed a Poisson process, then the distribution of $\beta$ was here generalized to a Poisson distribution whose probability mass function was defined by \eqref{eqn:joined_count_a}. And knowing the service times $t_{i}, t_{i+1},\ldots t_{z}$ were exponentially distributed, the density function of $T_{w}$ was expressed for using  \eqref{eqn:joined_count_b}:
\begin{subequations}
    \begin{equation}
       f(\beta) = \frac{\lambda^{\beta}e^{-\lambda}}{\beta!}, \quad \beta\in \mathbb{N} \\
       \label{eqn:joined_count_a}
    \end{equation}
    \begin{equation}
       f(T_{w}) = \mu e^{-\mu t_{w}}, \quad t_{w}\geq0
       \label{eqn:joined_count_b}
    \end{equation}
  \label{first:main}
\end{subequations}

\revise{\noindent}{}For a task waiting in $Q_{i}$ at position $k>L_{Q_{j}}$ therefore, the behaviour to move to $Q_{j}$ was here computed from the combined probabilities of new arrivals $\beta \leq \lambda$ being routed to $Q_{j}$ given $N \geq 1$ departures. Such that, when the jockeyed task landed at position $\tau_{i} = L_{Q_{j}}+\beta+1$, the expected waiting time $T_{w}$ at that position was less compared to staying at pose $k$. The second and third cases in \eqref{eqn:arrival_choices} were the trivial conditionals. 

\subsection{Expected jockeying frequency}

\revise{\noindent}{}The decision model here focused on the first case of \eqref{eqn:arrival_choices} to characterize for the frequency of jockeying $\xi$ given the eventualities arising from sensitivity variations and relations.  We encapsulated these relations in a function $g (T_{w}, \tau, N)$ defined by \eqref{eqn:quantity} such that the number of times jockeyed $\xi$ was a new random variable evolving from the existential inter-dependencies. 

\begin{equation}
 \xi = g(T_{w}, \tau,N)
   \label{eqn:quantity}
\end{equation}
However, it was only the expected mean waiting time $T_{w}$ that depended on the two independent events (the resultant position $\tau = L_{Q_{i\neq j}}+\beta + 1$ at which the job landed when the jockeying decision was taken and the number of $N\geq 1$ tasks that left either of the two buffers). And it was clear from Figure \ref{fig:gaussian} that $\tau$ was generally defined by \eqref{eqn:gaussian}. Under stable  system conditions and as earlier documented by \cite{algorith_approx}, the departures and arrival process followed a Poisson distribution, which had the implication that $N\leq \lambda$. 
Intuitively and as documented in \cite{danielgeorgio}'s findings, a higher number of departures $N\geq 1$ has a monotonic decreasing effect on the total waiting time $T_{w}$. It was therefore plausible to state that the frequency of jockeying $\xi$ was affected by the $N\geq 1$ and $\beta \leq \lambda$ jobs that joined the alternative queue.

\revise{\noindent}{}\textit{The Tower Property}:
\revise{\noindent}{}Given \eqref{eqn:gaussian},\eqref{eqn:joined_count_b} and deriving from the law of total probability (Tower rule), the theoretical notation for these dependencies were defined by  \eqref{eqn:cond_tw_tau} as the conditional probability on multiple occurrences in random variables $N$ and $\tau$.

\begin{equation}
  f_{T_{w}|N,\tau} (t_{w}|n,\tau) = \frac{f_{T_{w}|n,\tau} (t_{w},n,\tau)}{f_{\tau}(n|\tau)} = f_{T_{w}|\tau} \quad  \forall({ \tau, n})> 0 
  \label{eqn:cond_tw_tau}
\end{equation} 

\revise{\noindent}{}Deriving from Bayes' theorem, \eqref{eqn:joint_distr} arithmetically computed for the resultant conditional probability between $T_{w}$ and $\tau$.
\begin{equation}
f_{T_{w}|\tau} (t_{w}|\tau) =\int_{-\infty}^{\infty} \frac{f(t_{w}|\tau)f(t_{w})}{f(\tau)}  \quad  \forall({t_{w}, \tau})> 0
    \label{eqn:joint_distr}
\end{equation}

\revise{\noindent}{}From \eqref{eqn:eqn_tw} and \eqref{eqn:joint_distr}, \eqref{eqn:exp_tau_tw} expressed for the mean expected waiting time given these dependencies as:
\begin{equation}
  \mathbb{E}[T_{w} | \tau] = \int_{-\infty}^{\tau} t_{w}f_{T_{w}}(t_{w})\delta t_{w} = \sum_{i=1}^{m} \frac{t_{i,\tau}(1 - e^{-\mu t_{i,\tau} })}{m}
   \label{eqn:exp_tau_tw}
\end{equation}

\revise{\noindent}{}Here, the decision to jockey was synonymous to a Bernoulli process whose outcome was abstracted as a binary variable and \eqref{eqn:binary} characterized for this behavior.
\begin{equation}
   f_{\xi}(\xi) =
    \begin{cases}
     1 \qquad  & \text{if } T_{w,\tau} < T_{w,k} \\
     0 \qquad & \text{otherwise } 
    \end{cases}
   \label{eqn:binary}
\end{equation} 
This random variable was generalized to a binomial distribution whose probability mass function was given by \eqref{eqn:bernoulli}, definitive of the probability that exactly $\xi$ successes occurred in $d$ trials.
\begin{equation}
   P(\xi=\xi) = \binom{d}{\xi} p^{{\xi}} q^{d-{\xi}}
   \label{eqn:bernoulli}
\end{equation}
\textit{where p was the probability that the inequality $(H<G)$ holds true, $q = 1 - p$ and d the count of independent runs under preset arrival rate $\lambda$}.

\revise{\noindent}{}Denoting as $G$ the waiting time $T_{w,k}$ in the current queue (with service rate $\mu_{i}$) at position $k$, $H$ as the waiting time $T_{w,\tau}$ at position $\tau$ in the preferred queue (with service rate $\mu_{j}$), and given \eqref{eqn:joined_count_b}, the functional mapping to characterize for the inequalities between the two exponentially distributed random variables was computed using \eqref{eqn:inequality}.

\begin{equation}
  \begin{split}
   P(H < G ) =  \int_{0}^{\infty} P(H<G|G=g)f_{G}(g)dg \\
   = \int_{\tau}^{\infty} \int_{0}^{k} \mu_{j} e^{-\mu_{j}^{t_{w,\tau}}} \mu_{i} e^{-\mu_{i}^{t_{w,k}}} dt_{w,\tau} dt_{w,k}     
  \end{split}
  \label{eqn:inequality}
\end{equation} 
And resolving for \eqref{eqn:inequality} reduced to:
\begin{equation}
  P(H < G ) = \frac{\mu_{j}}{\mu_{j} + \mu_{i}} 
    \label{eqn:soln_inequal}
\end{equation}

\eqref{eqn:bayes_2} followed from the application of Bayes' theorem such that our waiting time model encapsulated the dependencies as the conditional probability between the jockeying frequency and the comparison between the waiting times.
\begin{equation}
   \begin{split}    
      F_{\xi=\xi|H<G}(\xi) = \frac{P((\{\xi=\xi\}) \bigcap P(H < G))}{P(H<G))}
   \end{split}
   \label{eqn:bayes_2}
\end{equation}
 
Computation for the average number of times workload was migrated then evolved from theoretic definitions (Law of Iterated Expectations and sub-$\sigma$ algebra) of the conditional expectation  representational of the relations between the dependent variables. Knowing that $\xi$ obeyed a binomial distribution $(d,p)$ with $d\in \mathbb{N}$ and  $p \in [0,1]$, then the expectation of $\xi \sim B(d,p)$ was formulated by \eqref{eqn:num_time}.
\begin{equation}  
      \mathbb{E}[\xi] = d*P(H<G)
  \label{eqn:num_time}
\end{equation}

And the analytical expression from  \eqref{eqn:num_time} took the form of \eqref{eqn:exp_final}. 
\begin{equation}
     \mathbb{E}[\xi] = \frac{d\mu_{j}}{\mu_{j}+\mu_{i}}
  \label{eqn:exp_final}
\end{equation}

\subsection{Assumptions during the experiment}
\begin{itemize}
   \item Admission control: Join the shorter queue - a widely adopted strategy for new entities joining the buffers. We however agree with \cite{shorter_queue,dehghanian2016} who argued against this being an optimal approach for communication networks given the dynamics in the underlying infrastructure environment. 
   \item Jockeying control: Jobs that had been waiting in either queues for the longest time were the first candidates for switching queues. At the time of new tasks joining the queues, the candidates were shuffled with the new arrivals to orchestrate a mixed scenario of resource competitiveness introduced by the probability that there were a portion $\beta \leq \lambda$ new arrivals joining the queue at a time $t+1$ when the task was rationally switched. 
 
\end{itemize}

\section{Empirical and Numerical Results}
Table \ref{tab:one} was a catalogue of the numerical computation of the queue size conditional probabilities as evaluated from \eqref{eqn:finalbayes} under the assumption that a single departure occurred at any given iteration. 
\begin{table}[htbp]
  \caption{With the parameters $Z = 5$, $\sigma = 1$,  $N =1 $ departures.} 
  \label{tab:one}
  \begin{tabular}{p{1.2cm} p{1.2cm} p{0.8cm} p{0.8cm} p{1cm} p{0.8cm} } 
    \toprule
    Len ($Q_{1}$) & Len ($Q_{2}$) & $\tau_{1}$ & $\tau_{2}$ & $P^{\beta}_{Q_{i,j}}$ & $\beta$\\
    \midrule
    5 & 2 & 5 & 7 & 0.49996 & 2.4998 \\
    2 & 10 &  7 & 10 & 0.83999 & 4.199995\\
    13 & 12 & 13 & 17 & 0.5 & 2.5 \\
    15 & 20 & 20 & 20 & 1.00 & 5.0\\
    1 & 4 & 6 & 4 & 0.5 & 2.5\\
  \bottomrule
\end{tabular}
\end{table}

\begin{figure*}[ht]
  \centering
  \includegraphics[width=0.95\linewidth]{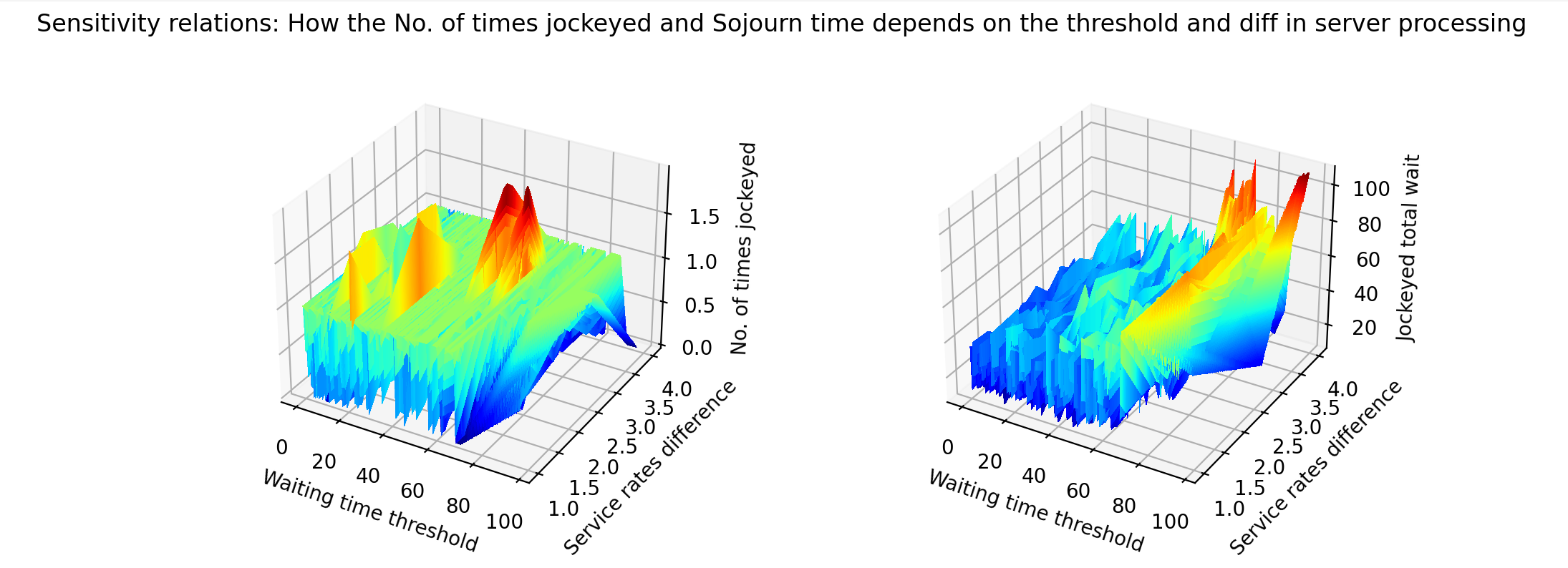}
  \caption{\footnotesize Sensitivity to variations in setup parameters: The figure on the left illustrated how combined variations in the processing rates of either queues affected the number of times the job was moved around. The graphic on the right in contrast captured the influence of these variations on the total time until service completion of the jockeyed job. It was observable that the differences in waiting times at the current position versus jockeyed to position had negligible effect on either measured descriptors.}
  \label{fig:3D_diff_count_threshold}
\end{figure*}

\begin{figure}[ht]
  \includegraphics[width=\linewidth]{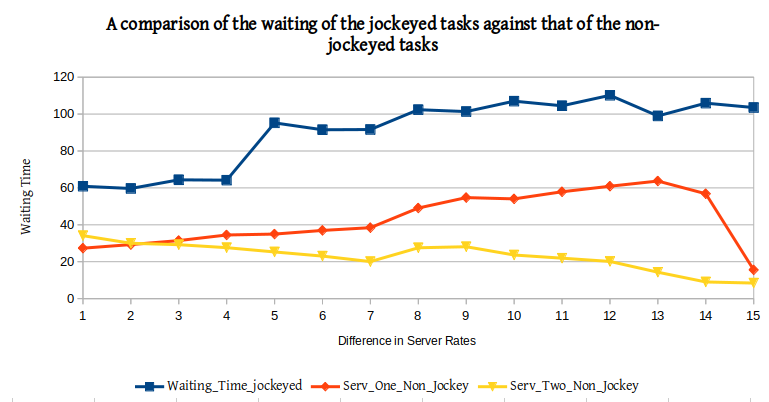} 
  \caption{\footnotesize A two dimensional extract that depicted the relationship between the difference in the capacities of the two buffers and the total time a jockeyed job consequently spent in the system against the jobs that were never migrated. }
  \label{fig:bi_variate}
\end{figure}

\begin{figure}[ht]
  \includegraphics[width=\linewidth]{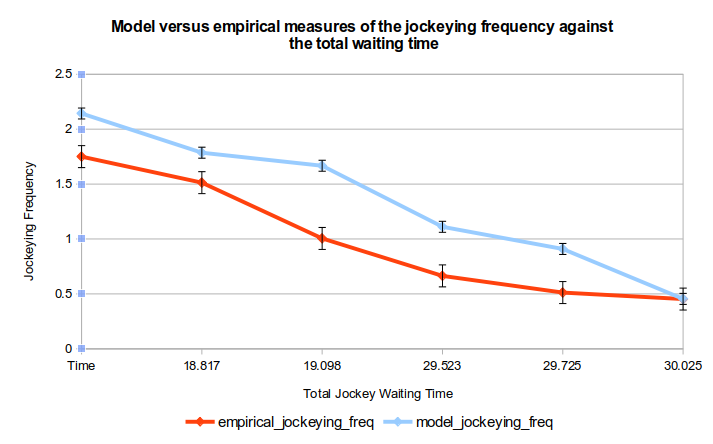}
  \caption{\footnotesize The jockeying customer consequently spent less time in the system. This profile augured with earlier studies to underline benefits of the behavior.} 
  \label{fig:frequency}
\end{figure}

Figure \ref{fig:3D_diff_count_threshold} was a depiction of the association between the deviations in the waiting times, the difference between the processing rates of the two buffers and what influence this heterogeneity in buffer configuration had on the number of times a task could traverse the system. The orientation of total waiting time profiles of the jockeyed jobs was also revealed. Figuratively, the dependencies were unravelled further by taking into account bi-dimensional comparisons between different descriptors as illustrated in Figure \ref{fig:bi_variate} and \ref{fig:frequency}. 
Figure \ref{fig:bi_variate} was affirmative of the analogy that the faster a queue was at processing tasks, the higher the intensity and this evolved into a scenario where one queue was under-utilized since tasks always sought to minimize the time to service completion at the expense of optimal system performance as a whole. The revelation was logical since to minimize the time spent in the system, the tasks always jockeyed to the faster queue, albeit the switch being a worse decision.  

Numerical evaluations for the modeled buffer descriptors were tabulated in Table \ref{tab:two} summarizing approximated measures on the jockeying frequency $\xi^{'}$ from the model as compared with those resulting from the simulated $\xi$. 
\begin{table}[htbp]
  \caption{Averaged measures from the simulation at different rates of the arrival and service rates of two queues with one queue always processing faster than the other $(\mu_{i} > \mu_{j})$. Computations for the model predicted jockeying frequencies $\xi^{'}$ assumed $d = 5$.} 
  \label{tab:two} 
  \begin{tabular}{p{0.6cm}p{0.7cm} p{0.7cm}  p{1.0cm} p{1.0cm} p{0.7cm} p{0.7cm} } 
    \toprule
    $\lambda$ & $\mu_{i}$ & $\mu_{j}$ & $T_{w,k}$ & $T_{w,\tau}$  & $\xi$ & $\xi^{'}$ \\ 
    \midrule
     7 & 4.0 & 3.0  & 18.653  & 18.817 & 1.75 & 2.1428 \\  
     7 & 4.5 & 2.5  & 19.098  & 17.487 & 1.51 & 1.7857 \\  
     9 & 6.0 & 3.0  & 29.523 & 26.694 & 1.005 &  1.667  \\ 
     9 & 7.0 & 2.0  & 29.725 & 25.740 & 0.664 & 1.1111 \\  
     11 & 9.0 & 2.0  & 30.025 & 25.274 & 0.512 & 0.909 \\  
     11 & 10.0 & 1.0  & 30.125 & 24.43 & 0.453 & 0.4545 \\  
  \bottomrule
\end{tabular}
\end{table}

\revise{\noindent}{}Figure \ref{fig:frequency} revealed an orientation akin to studies postulating that jobs that switched from one buffer to another spent less time in the system and this time decreased as the frequency of the behavior increased.
This was reflective of the benefits of jockeying and how the frequency of this behavior affected the expected waiting time in the system \cite{Gen.Solutions, Bi0bjective, ravid_2021}. These results gravitate with propositions from earlier analytic studies about the jockeying frequency of impatient customers, where the number of times $\xi$ that buffers were switched relative to the difference between the sizes of the buffers as the jockeying threshold was documented \cite{tarabia,zhao_grassman}. However, it could be argued that this proposition might not be valid when the accumulated costs are factored into the evaluation of the benefits. 

\section{Discussion and Future Work}
Numerous studies model for the jockeying behavior statistically where the behaviour is influenced by the difference between the length of the queues exceeding the preset limit \cite{koenigsberg,zhao_grassman}. Additionally, the stochastic studies adopt centralized control mechanisms for this behavior in a multi-queue setup. In contrast, our findings pioneer work in the class of behavioral models that seek to characterize for the impatient tenant's preferences. We lay the groundwork for efforts towards adoption of decentralized decision making such that, the decision to jockey is premised on measures about the tenant's continuous sensitivity to variations in queue parameters.
Our Monte Carlo experiments are based on our open source platform\protect\footnote{ \protect\url{https://github.com/anthonyKiggundu/Queuing-Theory}} that helps in first identifying correlative patterns within the buffer descriptors and extraction of proportionality of influence as dependent and independent variables. The selected descriptors then aggregate to guide the customer's decision to switch resource pools. The tooling eases the behavioral modeling of different types of queuing systems (like \ac{M/G/C}, \ac{G/G/C}) with configurable parameters. 
Having assessed the sensitivity relations that guided the impatient customers' preferences from the simulation, our decision model adopted concepts from probability theory to formulate for the frequency of this impatience numerically.

For future experiments, we still see room for improvements in the admission control as this could be further optimized by exposing new entrants to hybrid metrics like task migration costs, holding or overhead costs and expected waiting time before joining the queues. This is so as to ascertain how the degree of such knowledge influences the overall performance of the system given permitted jockeying behavior. Specific to \ac{6G} communication systems, realistic future setups could define the jockeying threshold based on costs and discounts or slice performance metrics (subscription costs, \ac{QoS} etc from the core or \ac{RAN} network functions (e.g. \ac{NSSF}, \ac{NWDAF}) in deviation from existing approaches that adopt buffer size differences 

Hypothetically, there should exist a limit on the number of times tasks are jockeyed such that under certain system conditions, the switching does not bring value anymore, thence counter-productive. It could be plausible to bound this impatience behavior then on the number of buffers $C$, that is $\xi \leq C$. However, the uncertainty introduced by the randomness makes the computation of this bound a stochastic optimization problem that would require further exploration of the statistical significance and correlational measures to unravel the magnitude of influence imposed by the various system descriptors. Consequently, quantitative measures could be tagged to the maximum number of times switching buffers in complex systems occurs and its benefits.

\revise{}{Finally, to leverage the benefits of the impatience behavior, performance benchmarking of the communication overhead introduced by broadcasting up-to-date queue status information is still an open discussion. Noteworthy too is how much value is pegged to this information \cite{Huang} or which queue descriptor combinations guide the decision making process best.}
\section*{Acknowledgment}
This work is supported by the \emph{German Federal Ministry of Education and Research (BMBF)} within the project Open6GHub under grant numbers 16KISK003K and 16KISK004. B. Han (bin.han@rptu.de) is the corresponding author.

\todo{I removed the acronym list section by adding the option [nolist] to the acronym package, since the list is not really required in this 6-page paper.}
\begin{acronym}[HRTEM]
  \acro{QoS}{Quality of Service}
  \acro{QoE}{Quality of Experience}
  \acro{NSSF}{Network Slice Selection Function}
  \acro{SMF}{Session Management Function}
  \acro{SDN}{Software Defined Networks}
  \acro{3GPP}{Third Generation Partnership Project}
  \acro{FCFS}{First Come First Server}
  \acro{LCFS}{Last Come First Serve}
  \acro{MEC}{Multi-access Edge Computing}
  \acro{5G}{Fifth Generation}
  \acro{6G}{Sixth Generation}
  \acro{M/M/C}{Markovian/ Markovian/ number of queues}
  \acro{M/G/C}{Markovian/General/ number of queues}
  \acro{G/G/C}{General/ General/ number of queues}
  \acro{RAN}{Radio Access Network}
  \acro{O-RAN}{Open Radio Access Network}
  \acro{DU}{Distributed Unit}
  \acro{RU}{Radio Unit}
  \acro{CU}{Centralized Unit}
  \acro{SIRO}{Serve In Random Order}
  \acro{NWDAF}{Network Data Analytics Function}
\end{acronym}
\bibliography{bibtex/bib/model.bib}{}
\bibliographystyle{IEEEtran}









\end{document}